\documentclass[12pt,a4paper]{article}
\usepackage[T1]{fontenc}
\usepackage{graphics}

\makeatletter

\newcommand{\LyX}{L\kern-.1667em\lower.25em\hbox{Y}\kern-.125emX\spacefactor1000}
\let\SF@@footnote\footnote
\def\footnote{\ifx\protect\@typeset@protect
    \expandafter\SF@@footnote
  \else
    \expandafter\SF@gobble@opt
  \fi
}
\expandafter\def\csname SF@gobble@opt \endcsname{\@ifnextchar[%]
  \SF@gobble@twobracket
  \@gobble
}
\edef\SF@gobble@opt{\noexpand\protect
  \expandafter\noexpand\csname SF@gobble@opt \endcsname}
\def\SF@gobble@twobracket[#1]#2{}

\usepackage{multicol}
\renewcommand{\vec}{\mathbf}
\makeatother

\begin{document}

\title{Theory of the Spin Reorientation Transition of Ultra-Thin Ferromagnetic Films}

\author{A.~Hucht and K.~D.~Usadel\\
\emph{\small Theoretische Tieftemperaturphysik, Gerhard-Mercator-Universität,
}\\
\emph{\small 47048 Duisburg, Germany}\small }

\date{{}}

\maketitle
%Version 1.5, Sep 23, 1998

\begin{abstract}
The reorientation transition of the magnetization of ferromagnetic films is
studied on a microscopic basis within Heisenberg spin models. Analytic expressions
for the temperature dependent anisotropy are derived from which it is seen that
the reduced magnetization in the film surface at finite temperatures plays a
crucial role for this transition. Detailed phase diagrams in the temperature-thickness
plane are calculated.
\end{abstract}
\if0\vspace{2cm}Keywords: reorientation transition, thin films, Heisenberg model.
\\
Contact author: 

A. Hucht, 

Theoretische Tieftemperaturphysik, 

Gerhard-Mercator-Universität Duisburg, 

D-47048 Duisburg 

Fax: x49-203-379-2965 

Email: fred@thp.Uni-Duisburg.DE

\newpage\renewcommand{\baselinestretch}{1.51}\small\normalsize
\fi

\markright{\rm \emph{Journal of Magn. and Magn. Mater.}, accepted for publication
(1998)}

\thispagestyle{myheadings}\pagestyle{myheadings}

The direction of the magnetization of thin ferromagnetic films depends on various
anisotropic energy contributions like surface anisotropy fields, dipole interaction,
and eventually anisotropy fields in the inner layers. These competing effects
may lead to a film thickness and temperature driven spin reorientation transition
(SRT) from an out-of-plane ordered state at low temperatures to an in-plane
ordered state at higher temperatures at appropriate chosen film thicknesses.
Experimentally, this transition has been studied in detail for various ultra-thin
magnetic films~\cite{allen1,pappas,qiu}. Recently, it was found by Farle~\textit{et~al.}~\cite{farle1}
that ultra-thin Ni-films grown on Cu(001) show an opposite behavior: the magnetization
is oriented in-plane for low temperatures and perpendicular at high temperatures.

It has been shown by various authors that the mechanism responsible for the
temperature driven transition can be understood within the framework of statistical
spin models. While Moschel~\textit{et~al.}~\cite{moschel1} showed that the
temperature dependence of the SRT is well described qualitatively within a quantum
mechanical mean field approach, most other authors focused on classical spin
models. Results obtained by extended Monte Carlo simulations~\cite{chui,hucht1}
as well as mean field calculations of classical Heisenberg models~\cite{hucht2,jb,hucht3},
although different in detail, agree in the sense that a temperature driven SRT
is obtained. Very recently, the perturbative approach~\cite{jb,hucht3} has
been extended to itinerant systems~\cite{herrmann}. Note, however, that due
to an unsufficient treatment of the spin-orbit induced anisotropy in this work
no SRT is possible for mono- and bilayers in disagreement with the above mentioned
theories and with experiments.  

While the SRT from a out-of-plane state at low temperatures to an in-plane state
at high temperature is found to be due to a competition of a positive surface
anisotropy and the dipole interaction the interesting new result for ultra-thin
Ni-films is argued~\cite{farle1} to have its origin in a stress-induced uniaxial
anisotropy energy in the inner layers with its easy axis perpendicular to the
film. This anisotropy is in competition with the dipole interaction and a negative
surface anisotropy. Theory confirms this picture~\cite{jb,hucht3}.  

In this paper a systematic investigation of the parameter and thickness dependence
of the reorientation transition and in particular a calculation of the corresponding
phase diagrams is presented. The calculations are done in the framework of a
classical ferromagnetic Heisenberg model consisting of \( L \) two-dimensional
layers on a simple cubic (001) or face centered cubic (001) lattice. The Hamiltonian
reads 
\begin{equation}
\label{hami}
\mathcal{H}=-\frac{J}{2}\sum _{\langle ij\rangle }\vec{s}_{i}\cdot \vec{s}_{j}-\sum _{i}D_{\lambda _{i}}(s_{i}^{z})^{2}+\frac{\omega }{2}\sum _{ij}r_{ij}^{-3}\vec{s}_{i}\cdot \vec{s}_{j}-3r_{ij}^{-5}(\vec{s}_{i}\cdot \vec{r}_{ij})(\vec{r}_{ij}\cdot \vec{s}_{j}),
\end{equation}
 where \( \vec{s}_{i}=(s_{i}^{x},s_{i}^{y},s_{i}^{z}) \) are spin vectors of
unit length at position \( \vec{r}_{i}=(r_{i}^{x},r_{i}^{y},r_{i}^{z}) \) in
layer \( \lambda _{i} \) and \( \vec{r}_{ij}=\vec{r}_{i}-\vec{r}_{j} \). \( J \)
is the nearest-neighbor exchange coupling constant, \( D_{\lambda } \) is the
uniaxial anisotropy which depends on the layer index \( \lambda =1\ldots L \),
and \( \omega =\mu _{0}\mu ^{2}/4\pi a^{3} \) is the strength of the long range
dipole interaction on a lattice with nearest-neighbor distance \( a \) (\( \mu _{0} \)
is the magnetic permeability and \( \mu  \) is the effective magnetic moment
of one spin). Note that only second order uniaxial anisotropies \( D_{\lambda } \)
enter the Hamiltonian (\ref{hami}). In our calculations we will restrict ourself
to the case that all anisotropies are equal to \( D_{v} \) except at \emph{one}
surface, where the value is \( D_{s} \), as this scenario is sufficient for
explaining the basic physics of the temperature driven reorientation transition.

The Hamiltonian (\ref{hami}) is handled in a molecular-field approximation,
the reader is referred to~\cite{hucht3} for further details. In the following
we assume translational invariance within the layers and furthermore that the
magnetization is homogeneous inside the film and only deviates at the surfaces,
which is justified by numerical simulations~\cite{hucht3}. Therefore we can
set \( \langle \vec{s}_{i}\rangle =\langle \vec{s}_{\lambda }\rangle  \) if
\( \vec{s}_{i} \) is a spin in a volume layer (\( \lambda =v \)) or in a surface
layer (\( \lambda =s \)). Now we must distinguish two cases:  

\begin{itemize}
\item For \( L<=2 \) we get a system with only one (surface) layer with the effective
interactions \( \tilde{x}_{ss}=x_{0}+2x_{1}(1-L_{s}^{-1}) \), where \( x \)
is either the coordination number, \( z \), or the dipole sum, \( \Phi  \),
respectively. In this case we have \( L_{s}=L \) and \( L_{v}=0 \). The effective
uniaxial anisotropy is \( \tilde{D}_{s}=D_{s}+(L_{s}-1)D_{v} \) and the Curie
temperature of the unperturbated system is \( T_{c}(L)/J=\tilde{z}_{ss}/3 \).
 
\item For \( L>2 \) we get a system with two layers with the effective interactions
\( \tilde{x}_{ss}=x_{0} \), \( \tilde{x}_{vs}=x_{1} \), \( \tilde{x}_{vv}=x_{0}+2x_{1}(1-L_{v}^{-1}) \),
and \( \tilde{x}_{sv}=2x_{1}L_{v}^{-1} \). Here we have \( L_{s}=2 \) and
\( L_{v}=L-2 \). Note that the coupling between the volume spin and the surface
spin is asymmetric. The effective uniaxial anisotropies are \( \tilde{D}_{s}=D_{s}+D_{v} \)
and \( \tilde{D}_{v}=L_{v}D_{v} \), respectively. The critical temperature
of the unperturbated system is given by 
\[
\frac{T_{c}(L)}{J}=\frac{z_{0}}{3}+\frac{z_{1}}{3}\left( 1-L_{v}^{-1}+\sqrt{1+L_{v}^{-2}}\right) .\]

\end{itemize}
In this approach we are not restricted to films with integer values of the thickness.
The number of next neighbors \( z_{\delta } \) and the constants \( \Phi _{\delta } \)
containing the effective dipole interaction between the layers for different
lattice types and orientations are listed in table~\ref{tab}.\begin{table}[ht]
{\centering \begin{tabular}{c|ccc|cr@{}lc}
lattice&
\( z_{0} \)&
\( z_{1} \)&
\( z_{\delta >1} \)&
\( \Phi _{0} \)&
\multicolumn{2}{c}{\( \Phi _{1} \) }&
\( \Phi _{\delta >1} \)\\
\hline 
sc(001)&
4&
1&
0&
9.034&
-0.&
327&
\( \mathcal{O}(e^{-2\pi (\delta -1)}) \)\\
fcc(001)&
4&
4&
0&
9.034&
1.&
429&
\( \mathcal{O}(e^{-\sqrt{2}\pi (\delta -1)}) \)\\
\end{tabular}\par}

\caption{Numerical constants \protect\( z_{\delta }\protect \) and \protect\( \Phi _{\delta }\protect \)
for sc(001) and fcc(001) lattices.}

\label{tab}
\end{table} 

A perturbation theory to the mean field Hamiltonian can be applied considering
the terms proportional to \( \omega  \) and \( D_{\lambda } \) as small perturbations
which is justified in view of the smallness of these parameters~\cite{jb,hucht3}.
In this framework we derived an analytical expression for the total anisotropy
\( K(T) \) of the system which is defined as the difference of the free energies
of the in-plane state and the out-of-plane state. \( K(T) \) depends on temperature
only through the absolute value of the subsystem magnetizations \( m_{\lambda }(T) \)
calculated with the unperturbed Hamiltonian and is given by 
\begin{equation}
\label{ktot}
K(T)=\sum _{\lambda }\tilde{D}_{\lambda }\left( 1-3T\frac{m_{\lambda }(T)}{h_{\lambda }(T)}\right) -\frac{3\omega }{4}\sum _{\lambda ,\lambda '}m_{\lambda }(T)\tilde{\Phi }_{\lambda \lambda '}L_{\lambda '}m_{\lambda '}(T).
\end{equation}
\( \vec{h}_{\lambda }=J\sum _{\lambda '}\tilde{z}_{\lambda \lambda '}\vec{m}_{\lambda '} \)
is the mean field in subsystem \( \lambda  \). If we neglect the narrow canted
phase, the reorientation temperature \( T_{r} \) is given by the condition
\( K(T_{r})=0 \).  

For the homogeneous case (\( L<=2 \)) the magnetization is approximately given
by 
\[
m^{2}(T)\approx \frac{1-T/T_{c}}{1-2T/5T_{c}}.\]
 This approximation deviates less than 1\% from the values obtained numerically.
From (\ref{ktot}) we then get 
\[
K(T)\approx \left( 1-\frac{T}{T_{c}}\right) \left[ \tilde{D}_{s}-\frac{3\omega L\tilde{\Phi }_{ss}}{4(1-2T/5T_{c})}\right] ,\]
 from which the reorientation temperature is determined as 
\[
\frac{T_{r}}{T_{c}}\approx \frac{5}{2}-\frac{15\omega }{8}\frac{\Phi _{0}L+2\Phi _{1}(L-1)}{D_{s}+(L-1)D_{v}}\]
 \begin{figure}[ht]
{\centering \resizebox*{12cm}{!}{\includegraphics{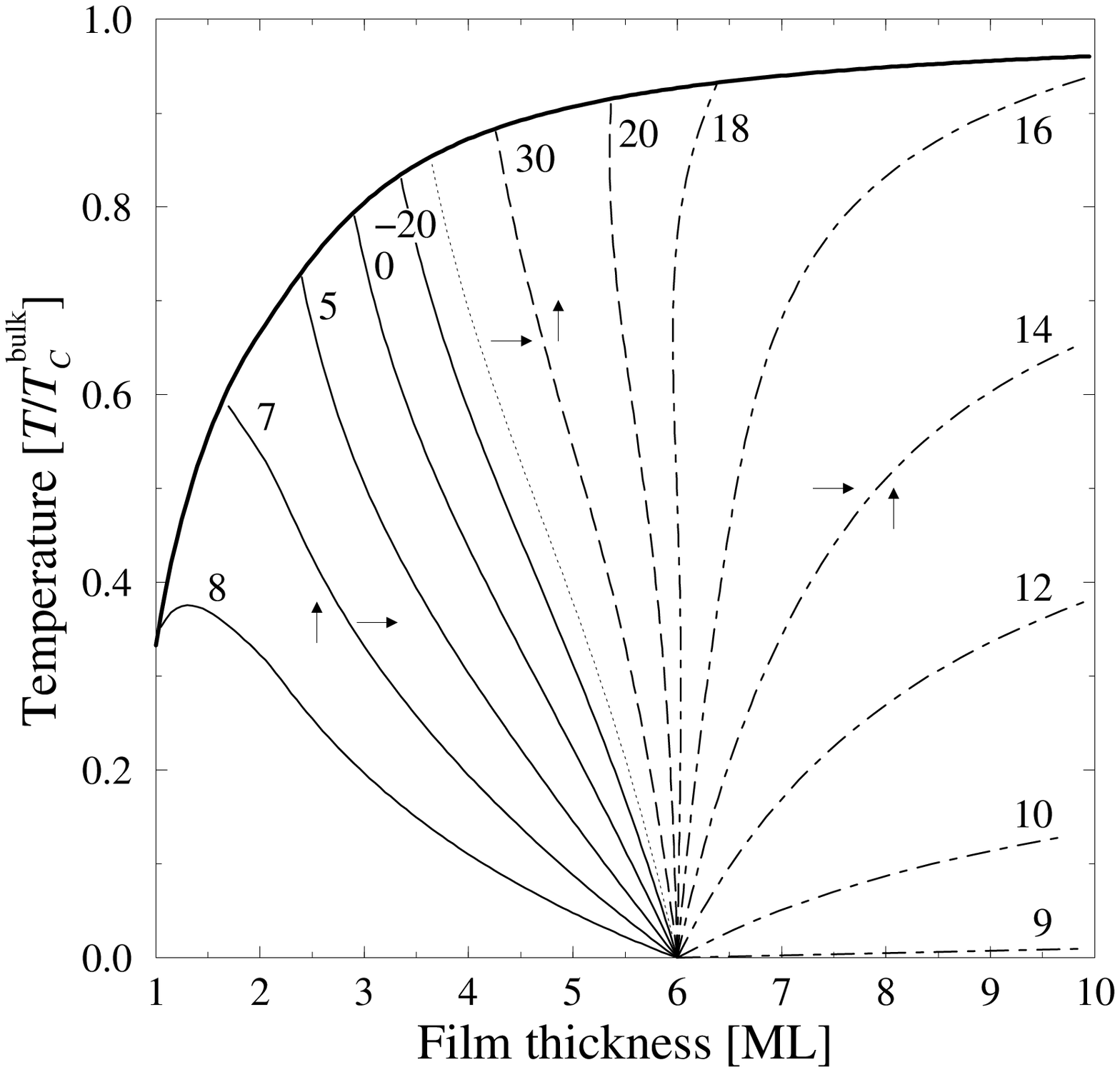}} \par}

\caption{Phase diagrams of the fcc system for several values of \protect\( D_{v}/\omega \protect \)
(Numbers). \protect\( D_{s}/\omega \protect \) is chosen so that in the ground
state the SRT occurs at \protect\( L_{r}(T=0)=6\protect \), i.e.~\protect\( D_{s}=51.37\omega -5D_{v}\protect \).
The thick solid line represents the Curie temperature \protect\( T_{c}\protect \)
of the film, the solid lines are normal (Fe-type) SRTs, while the dashed lines
are reversed (Ni-type) SRTs. The dash-dotted lines represent a third form of
SRT, where the reorientation temperature \protect\( T_{r}\protect \) grows
with film thickness and does not necessarily reach \protect\( T_{c}\protect \)
for finite values of \protect\( L\protect \). At the thin dotted line \protect\( D_{v}/\omega \rightarrow \pm \infty \protect \),
i.e. here the SRT occurs for \protect\( \omega =0\protect \).}

\label{fig_L6}
\end{figure}The phase diagram of the fcc(001) system calculated within this approach is
shown in figure~\ref{fig_L6}. One can distinguish \emph{three} different reorientation
schemes: 

\begin{itemize}
\item For uniaxial volume anisotropies 
\[
\frac{D_{v}}{\omega }<\frac{3}{4}(\Phi _{0}+2\Phi _{1})\approx 8.92\quad (\mathrm{fcc})\]
 we find a normal SRT from perpendicular to in-plane magnetization with increasing
temperature and film thickness (solid lines in figure \ref{fig_L6}). 
\item If \( D_{v} \) is large, we find a reversed SRT from in-plane to perpendicular
magnetization with increasing temperature and film thickness (dashed lines in
figure \ref{fig_L6}). 
\item In the intermediate range we obtain a third type of SRT where the magnetization
switches from perpendicular to in-plane direction with \emph{increasing} temperature,
but with \emph{decreasing} film thickness. (dash-dotted lines in figure \ref{fig_L6}). 
\end{itemize}
This work was supported by the Deutsche Forschungsgemeinschaft through Sonderforschungsbereich
166. One of us (A.H.) would like to thank M. Farle for fruitful discussions. 

%\newpage

%\end{multicols}%\newpage\vspace*{4cm}

%\newpage

\end{document}